 \documentclass[10pt, a4paper]{article}
 \usepackage{graphicx}                    % For eps figures, newer & more powerfull
 \usepackage{color}                       % For color text: \color command
 \usepackage{url}                         % For breaking URLs easily trough lines
\begin{document}

 %\begin{article}

%\begin{opening}

Title: Longitudinal structure of the photospheric magnetic field in Carrington system.

Authors: E. A. Gavryuseva (Institute for Nuclear Research RAS)

Comments: 9 pages, 4 Postscript figures

 \begin{abstract}
   The observations of the Sun have been performed over the years, even centuries-
   Whether are there active longitudes? If yes how stable are they?
   ne of the first
   The Wilcox Solar Observatory data taken over three cycles N 21, N 22, N 23 have been used 
   to reveal the longitudinal structure of the photospheric magnetic field.
   Mean over three cycles magnetic field distribution has been calculated in the North and 
   in the South hemispheres as well as at 30 levels of latitude from -75 to 75 degrees. 
   This study was performed using observations of the magnetic field taking into account its polarity or only intensity.
   The longitudinal structure of the magnetic field was calculated in the coordinate system rotating with Carrington rate.
   
   These structures were compared with a model of random longitudinal distribution of the magnetic field.
   Random character of the longitudinal structure was refused. 
   
   The results agree with the presence of two active meridians
   seen in different phenomena of solar activity at longitudes
   separated by 150-170 degrees in the Carrington coordinate system.
   \end{abstract}
 \vspace{1pc}

 %%%%%%%%%%%%%%%%%%%%%%%%%%%%%%%%%%%%%%%%%%%%%%%%%%%%%%%%%%%%%%%%
        \section{Introduction and data}
 Are there active longitudes on the Sun?

If there are, how stable are they? What is the lifetime of active longitudes?

If so, which latitudes contribute the most?

On the Sun there are various dynamic processes and hundreds of articles 
were written about the analysis of perturbations of solar characteristics, 
their distribution over the surface of the Sun and their temporal behavior 
(see for example Bumba et all., 1965, 1969, 1987 and references there)

Magnetic fields and the rotation of the Sun are the main factors which determine 
the variations and activity observed on the surface of the Sun.

To study these problems, it is necessary to analyze the series of data on 
the magnetic fields of the Sun measured at different latitudes during several solar cycles. 
Observations of the photospheric magnetic field were performed
by the Babcock solar magnetograph on WSO
  using the Zeeman splitting Fe I spectral line 
  (Scherrer et al., 1977; Hoeksema, 1984; 1985).
  The data sets have  5 degree steps in heliographic longitude.
  But each longitudinal value is a weighted average of the measurements within 
  55 degrees around central meridian. Along the latitude the grid of data have 30 levels
 with latitude $\theta$ is changing from   75.2 North to 75.2 South degrees.

    This paper is focused on study of the longitudinal structure
  of the  photospheric magnetic field over solar cycles NN21--23.

 %%%%%%%%%%%%%%%%%%%%%%%%%%%%%%%%%%%%%%%%%%%%%%%%%%%%%%%%%%%%%%%%
 \section{Longitudinal structure of the Solar Magnetic Field}
  The presence of active longitudes could contribute
  to recurrent perturbations of  the geomagnetosphere.
  Longitudinal activity was a subject of detailed studies   with
  different tracers in the rigidly rotating coordinate system.
                      
  Rotation rate of the SMF depends on the level
  where it is originated  and on the rotation
  of the plasma above this level.
                      
  Two approaches could be used for the analysis of
  the longitudinal structure of the magnetic field.
                      
  If the SMF activity  is originated from a deep rigidly rotating
  level then the differential rotation of the upper layers would 
  influence the original distribution of the  activity.
  In this case the  longitudinal structure reconstruction
  should be performed taking into account the rotational rate
  of the photospheric plasma.
 Since we know that this  rotational rate is changing in time
 the coordinate system should follow this variation too.
 The results of this reconstruction can be found in the paper
 of  Gavryuseva and Godoli  (2006);  Gavryuseva, (2006c, 2006e, 2008b).
                      
 If the  level from which the SMF is originated is not too deep
 under the photospheric level  and the SMF is rigidly rotating
 then the SMF longitudinal structure reconstruction
 should be performed  using the SMF as they are
 observed.            
 In this paper we concentrate on the global
 structure of the SMF in the coordinate system rigidly  rotating with the 
 Carrington period equal to 27.2753 days and  we present here
 the results related to the longitudinal structure of the
 photospheric field in the same system.
                      
 It is plotted in  Fig. 1  the distribution of the magnetic field on the
 solar surface along the latitudes  from -75 to 75 degrees
 and the longitudes from 0 to 360 degrees mean over 260 Carrington
 rotations (covering two solar cycles No 21 and No 22 since CR No 1642).
 The yellow (blue) colors correspond to high positive
 (up to 170 micro Tesla)
 and negative (up to -125 micro Tesla) values
 of this SMF mean latitudinal-longitudinal
 distribution located around  active latitudes
 and in pre-equatorial zones.
 The contours correspond to the
 $0, \pm 50, \pm 100$ micro Tesla.
 The averaging over all latitudes
 gives the mean longitudinal distribution,
 which is plotted in  Fig. 2 by continuous line.
 The deviation from zero level of this
 longitudinal distribution  is varying from about -30
 to 40 micro Tesla.  
 
   \section{Random magnetic field distibution}                    
 This  latitudinal-longitudinal distribution should be compared with the ones
 corresponding to models with random SMF longitudinal distributions
 shown in Fig. 3 by dots.
 (The adequate model of the random SMF longitudinal distribution should have
 all the  main characteristics of the solar activity
 (discussed in Gavryuseva (2018a, 2018b)).
 The models of the random SMF longitudinal distribution models were
 calculated by randomization of the real SMF
 for each Carrington Rotation at each latitude.
 With this procedure all the important characteristics 
 (latitudinal distribution, activity cycles, etc.) 
 are present in the model. The only difference is
 the random character of the SMF longitudinal distribution.
 Longitudinal mean of the random SMF over two cycles of activity
 No 21 and No 22 was  calculated.
 Both longitudinal distributions
 mean over 260 CR since 1642 CR are plotted in Fig. 3
 for the real SMF by continuous line and
 for ten models of the random SMF longitudinal distributions by dots. 
 The amplitude of the longitudinal distribution of the real SMF
 is 7-10 time higher
 than the amplitude of the models of the random SMF.
 This confirms  that the real SMF longitudinal distribution
 can not be described by the models of the random SMF distribution.

  \section{Longitudinal structure of the Magnetic Field intensity}
 The longitudinal activity is characterized by the amplitude of the
 magnetic field and not only by its polarity which is changing in time.
 To prevent the reduction (or even annihilation) of
 the mean magnetic field over 28-year  interval due to the
 various inversions of polarity,
 the longitudinal distribution of the absolute value of
 the SMF averaged  over all the latitudes and over 260 CR rotations
 (longitudinal distribution of the magnetic field intensity).
 has been computed and plotted in Fig. 4.
 This curve could be compared with
 the  distributions corresponding
 to the ten models of the random longitudinal distribution
 of the SMF intensity shown by dots on the same plot.
 The random longitudinal SMF distribution is 6.5 times lower
 than the longitudinal distribution of the absolute values 
 of the measured SMF.
                      
 The longitudinal distribution of the intensity
 of the solar magnetic field
 shows that there are two active longitudes
 around 10 and 220 degrees. This result
 is in a good agreement with the conclusions
 of other studies related to the longitudinal distribution
 of different characteristics of solar activity
 (Vitinsky, 1969;
  Bai, 1988;    
  Mordvinov and Plyusnina, 2000;
  Ivanov et al., 2001).
                      
  These results exclude the possibility of the interpretation
  of the longitudinal distribution of the photospheric magnetic
  field by the models of fully random perturbations or, 
  that is more interesting, by the models of fully stochastic 
  origin of the solar magnetic field longitudinal distribution 
  on the time scale significantly shorter than duration of solar cycle
  and on the spacial scale less than  solar radius.

  An additional recent study confirmed the existence of long living
  magnetic field meridional structures (longer than 20 years)
  originated from the bottom of the convective zone and rotating
  rigidly with the period corresponding to the latitude of 55 degrees about
  Gavryuseva, (2008a,b).
 
 \section{Summary}
    Longitudinal structure that is quasi-stable over several years
   has been found in the Carrington coordinate system in the
   large scale magnetic field observations
   and compared with models of a random longitudinal SMF distribution.
   
  This magnetic field topology, highly-organized over the solar surface
  and over time,   must  be considered as a basic structure with
  a major influence  on the solar corona and solar wind propagation,
  and is fundamental for the understanding
  of the heliospheric structure and
  for the prediction of the magnetospheric perturbations.

  %%%%%%%%%%%%%%%%%%%%%%%%%%%%%%%%%%%%%%%%%%%%%%%%%%%%%%%%%%%%%%%%   
 %%%%%%%%%%%%%%%%%%%%%%%%%%%%%%%%%%%%%%%%%%%%%%%%%%%%%%%%%%%%%%%% 
%%%%% \begin{acks}
\section*{Acknowledgments}
   I thank very much the WSO team for their great efforts
   in the measurements of the photospheric field.
   Thanks a lot to Prof. L. Paterno and Dr. E. Tikhomolov
   for precise and profitable advises and
   help in preparation of this paper.
   I am very grateful to Prof. B.T. Draine for his help
   in the revision  of this paper.
%%%%% \end{acks}

 %%%%%%%%%%%%%%%%%%%%%%%%%%%%%%%%%%%%%%%%%%%%%%%%%%%%%%%%%%%%%%%%

 \newpage
 \clearpage
 %%%%%%%%%%%%%%%%%%%%%%%%%%%%%%%%%%%%%%%%%%%%%%%%%%%%%%%%%%%%%%%%%%%%%%%%%%
 %%%%%%%%%%%%%%%%%%%%%%%%%%%%%%%%%%%%%%%%%%%%%%%%%%%%%%%%%%%%%%%%%%%%%%%%%%

%%%%%%%%%%%%%%%%%%%%%%%%%%%%%%%%%%%%%%%%%%%%%%%%%%%%%%%%%%%%%%%%%%%%%%%
%%%   14
 \begin{figure}
\centerline{
\includegraphics[width=39pc]
   {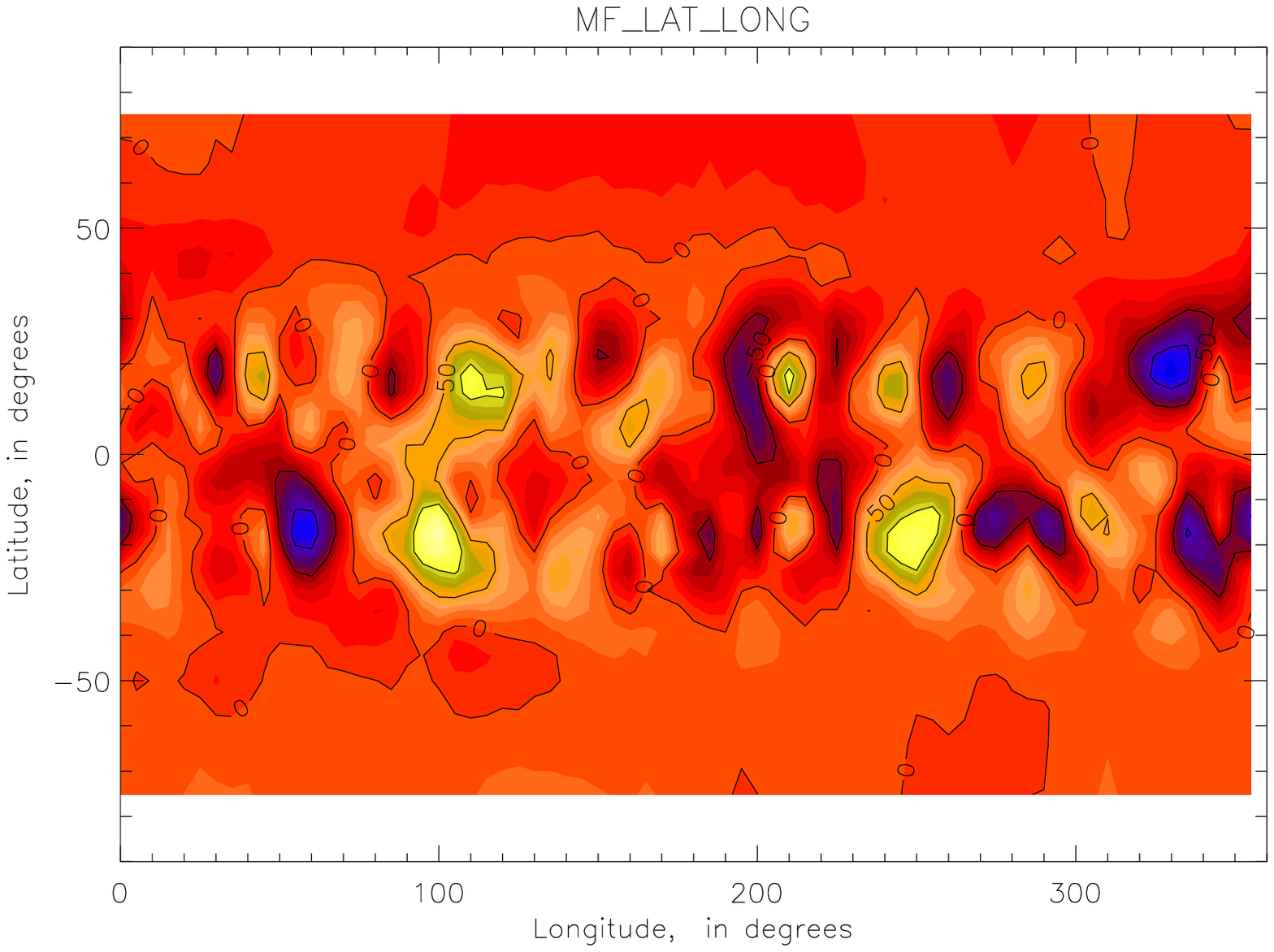}
}
\caption{
  Distribution in longitude and in latitude
  of the mean over 260 Carrington rotations
  real magnetic field on the solar surface.
  The yellow (blue) colors correspond to high positive
  and negative  values  of this SMF distribution.
  The contours correspond to the $0, \pm 50, \pm 100$ micro Tesla.
 }
  \end{figure}
%%%%%%%%%%%%%%%%%%%%%%%%%%%%%%%%%%%%%%%%%%%%%%%%%%%%%%%%%%%%%%%%%%%%%%%
%%%   15
 \begin{figure}
\centerline{
\includegraphics[width=39pc]
   {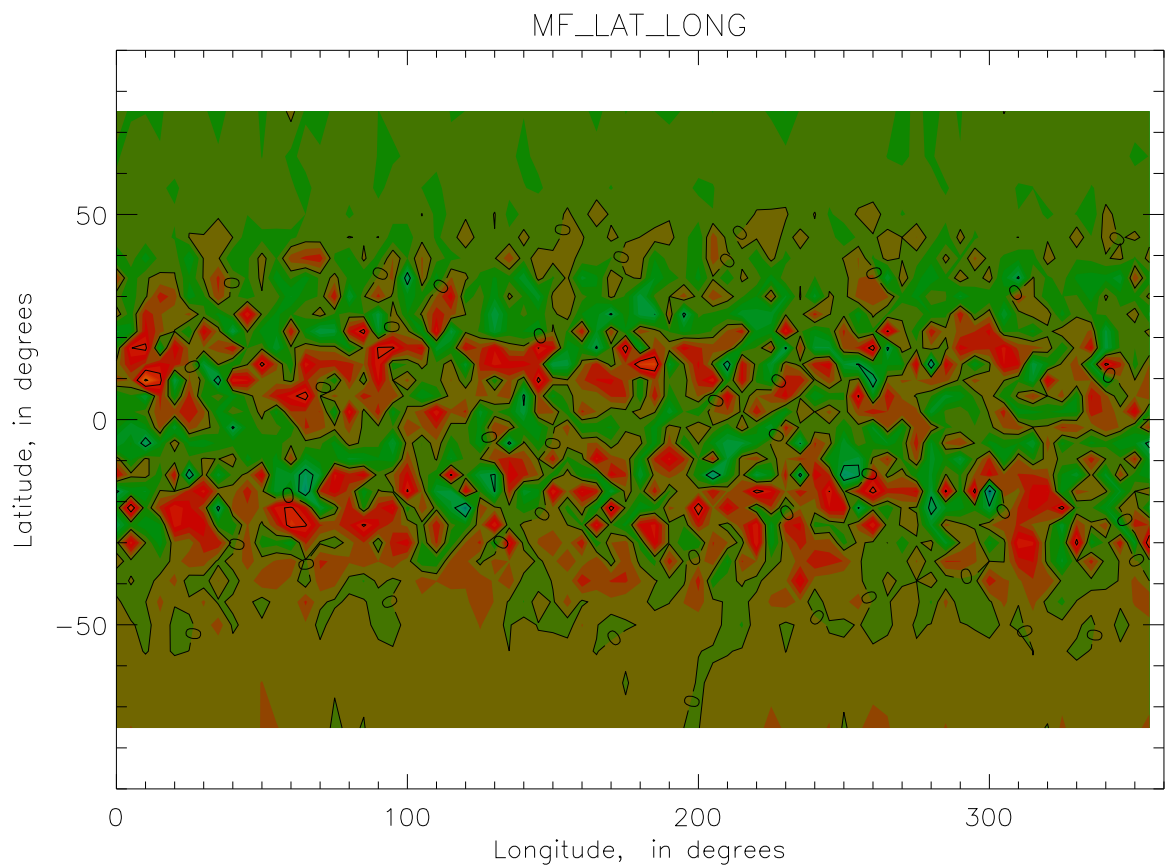}
%%%%%   {REFgavryuseva1_fig15.ps}
}
  \caption   
 {  
 Longitudinal distribution of the mean over all latitudes
 and averaged over two solar activity cycles (No 21 and No 22) 
 is shown by continuous line.
 The longitudinal distributions of ten models of the SMF randomly 
 distributed along the longitudes are plotted by dots.
 }
  \end{figure}
%%%%%%%%%%%%%%%%%%%%%%%%%%%%%%%%%%%%%%%%%%%%%%%%%%%%%%%%%%%%%%%%%%%%%%%
%%%   16
 \begin{figure}
\centerline{
 \includegraphics[width=39pc]
   {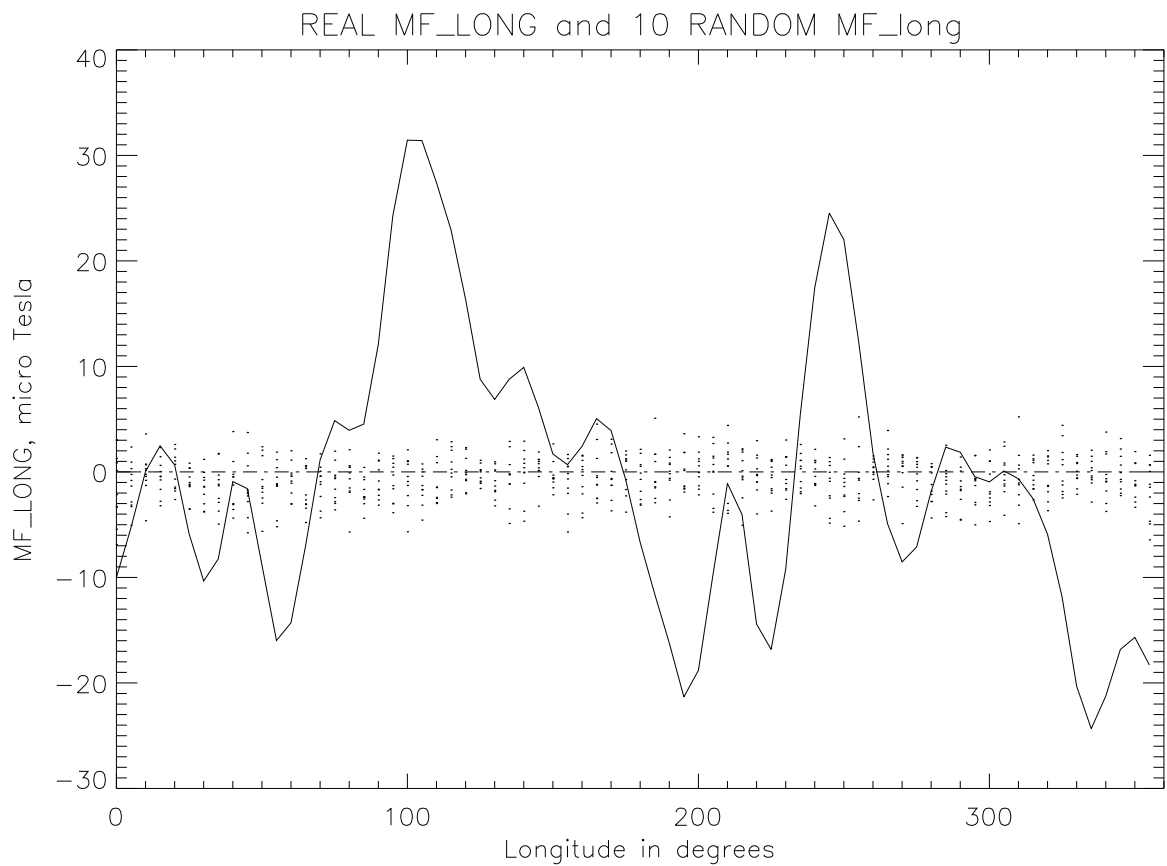}
%%%%%   {REFgavryuseva1_fig16.ps}
}
  \caption   
 {  
  Distribution in longitude and in latitude
  of the mean over 260 Carrington rotations
  magnetic field  on the solar surface
  for the model with random longitudinal SMF distribution.
  The contours correspond to the $0, \pm 50, \pm 100$ micro Tesla.
  }
  \end{figure}
%%%%%%%%%%%%%%%%%%%%%%%%%%%%%%%%%%%%%%%%%%%%%%%%%%%%%%%%%%%%%%%%%%%%%%%
%%%   17
 \begin{figure}
\centerline{
 \includegraphics[width=39pc]
   {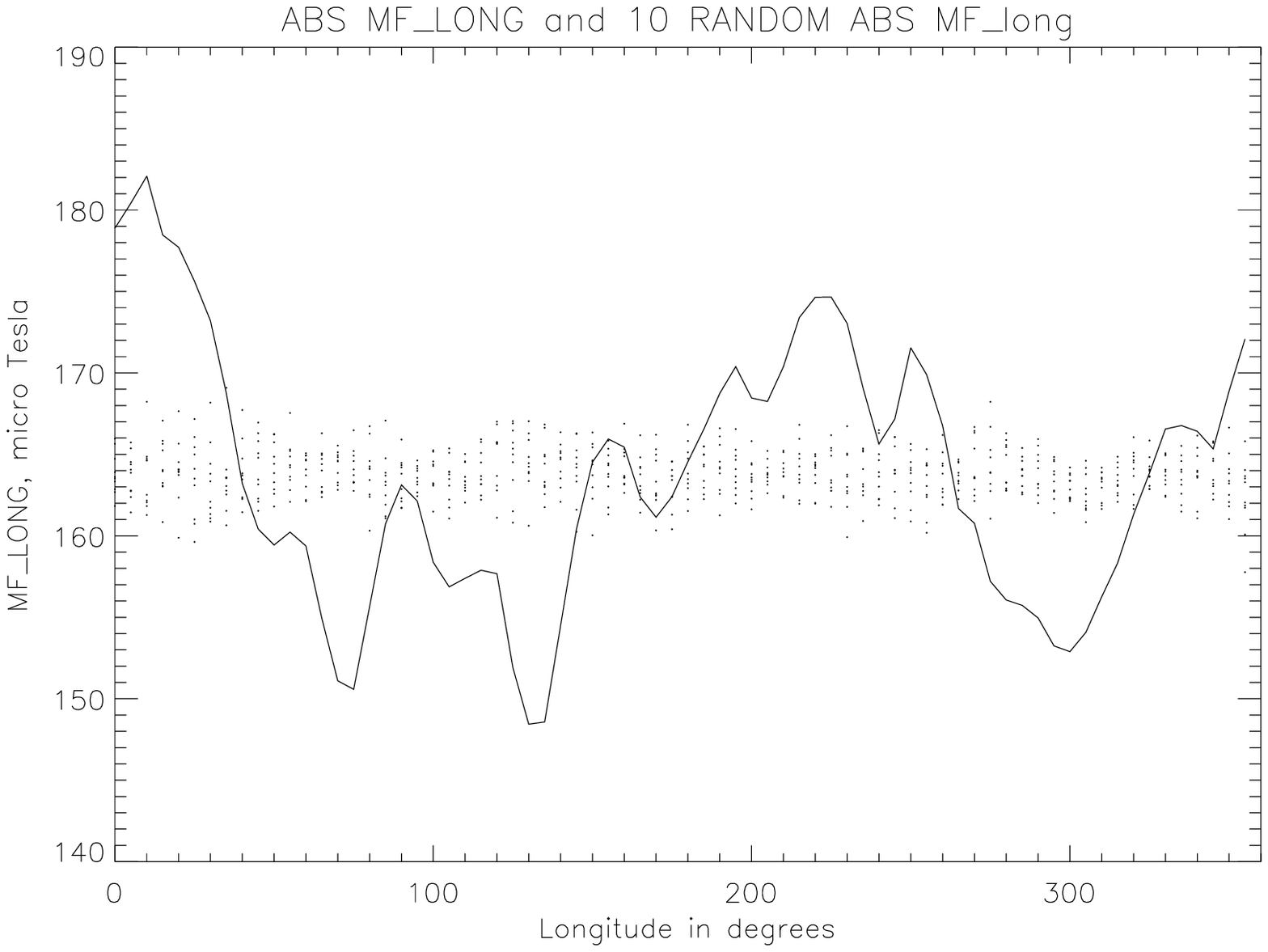}
}
 {  
 The  mean longitudinal distribution averaged over all latitudes
 and two solar activity cycles (No 21 and No 22) for the 
 photospheric magnetic field intensity is shown by  continuous line.
 The longitudinal distributions of ten models of the SMFI randomly  
 distributed along the longitudes are plotted by dots.
 }
  \end{figure}
 %%%%%%%%%%%%%%%%%%%%%%%%%%%%%%%%%%%%%%%%%%%%%%%%%%%%%%%%%%%%%%%%%%%%%%%%%
   %\end{article} 
   \end{document}